# Measuring Large Language Models Dependency: Validating the Arabic Version of the LLM-D12 Scale

Sameha AlShakhsi[1], Ala Yankouskaya[2], Magnus Liebherr[3], Raian Ali[1]

[1]College of Science and Engineering, Hamad Bin Khalifa University, Qatar
`{salshakhsi, raali2}@hbku.edu.qa`
[2]Department of Psychology, Bournemouth University, UK
`ayankouskaya@bournemouth.ac.uk`
[3]Department of Mechatronics, University Duisburg-Essen, Germany
`magnus.liebherr@uni-due.de`

**Abstract**. There is an urgent need for reliable, culturally validated instruments to assess psychological responses to AI in general and large language models (LLMs). This need is global issue, but it is especially urgent among Arabic-speaking populations, where AI and LLMs adoption is accelerating, yet psychometric tools remain limited. This study presents the first validation of the LLM-D12, a dual-dimensional scale assessing Instrumental and Relationship Dependency on LLMs, in an Arab sample. A total of 250 Arab participants completed the Arabic version of the LLM-D12. Confirmatory Factor Analysis confirms the original 2-factor structure of LLM-D12 with all items showing good loading of corresponding Instrumental and Relationship Dependency. The scale showed good to excellent internal reliability (Cronbach alpha is 0.90 for Total, 0.85 for Instrumental Dependency, and 0.90 for Relationship Dependency). External validation revealed that Instrumental Dependency was positively associated with AI acceptance and internet addiction, while Relationship Dependency was linked to lower need for cognition and greater trustworthiness of LLM, demonstrating sensitivity of this instrument to different use and personal factors. These findings confirm that Arabic LLM-D12 is a psychometrically sound, culturally appropriate instrument, offering a necessary tool for research, education, and policy concerning AI and LLMs engagement in Arab contexts.

**Keywords**: Large Language Models, Dependency, Relationship Dependency, Instrumental Dependency, Generative AI, Human-AI Interaction

## INTRODUCTION

Large Language Models (LLMs) are becoming increasingly embedded in everyday life, with users across the globe turning to them not only for factual information, but also for advice, emotional support, and decision-making. In the Arab world, this trend is particularly evident. By 2024, 88% of CEOs in Gulf Cooperation Council (GCC) countries reported adopting generative AI [1], signalling rapid institutional integration. Across the region, LLMs are now used for a wide range of tasks, from translation and administrative tasks to more socially complex roles such as educational assistance, health-related queries, and even conversational companionship. Their widespread uptake indicates not only a regional appetite for digital innovation but also the potential for deepening relational bonds between users and machines.

The growing reliance on Large Language Models (LLMs) such as OpenAI's ChatGPT, Google's Gemini, and DeepSeek reveals several features that may facilitate habitual and potentially addictive patterns of use. Instrumentally, users benefit from structured problem-solving, memory support, and cognitive offloading [2][3][4][5], but repeated delegation of reasoning and decision-making may diminish critical thinking, creativity, and independent judgement over time [6][7][8][9]. This reliance is reinforced by the immediacy and personalisation of responses, which offer cognitively and emotionally rewarding interactions [10]. At the same time, the conversational tone, emotional responsiveness, and adaptive feedback of LLMs simulate human-like companionship, encouraging self-disclosure and the formation of parasocial bonds [11][12][13][14]. Such interactions, while initially comforting, may risk substituting real-world social interactions over time [10][15]. As human reliance on LLMs for both task support and emotional connection continues to grow, there is a pressing need to measure this emerging form of dependency in order to better understand its nature, scope, and implications.

Despite increasing recognition of LLM dependency as a potential concern [10], the availability of validated tools for its assessment remains limited. The Problematic ChatGPT Use Scale (PCUS) was one of the first instruments designed to assess problematic ChatGPT use, originally developed in a Taiwanese context [16], and later adapted for Turkish users [17]. The Problematic Use of Conversational Artificial Intelligence Scale (PUCAI-6) is a brief six-item measure adapted from the Bergen Social Media Addiction Scale, which captures problematic use of conversational AI in relation to psychological factors such as social anxiety, loneliness, and rumination [18]. Other available tools include a three-item scale for generative AI addiction [19] and the eight-item AI Chatbot Dependence Scale (AICD) [20], an adaptation of smartphone addiction frameworks to assess AI-related dependency [21]. A more conceptually differentiated model distinguishes between functional and existential forms of LLM dependency, drawing on theories of trust and addiction to explain users' task-related reliance versus deeper emotional attachment [22]. However, these measures primarily adapt general behavioural addiction frameworks to the LLM context, which may not fully capture the unique dynamics of human-LLM interaction [23].

A separate line of research has introduced the Large Language Model Dependency Scale (LLM-D12), a conceptually distinct instrument developed to capture LLM-specific patterns of problematic reliance and dependency [23]. LLM-D12 does not rely on the symptoms typically used to operationalize other forms of behavioural addiction, such as social media or gaming disorder, and consider the special nature of LLMs purposes of use and interactions. LLM-D12 is a 12-item measure comprising two subscales: Instrumental Dependency (6 items) and Relationship Dependency (6 items). Instrumental Dependency reflects users' over-reliance on LLMs for completing tasks, solving problems, and making decisions, besides users' tendency to feel more rewarded and take more job in the presence of LLMs. Relationship Dependency captures affective tendencies such as emotional support seeking, and the formation of perceived social bonds and companionship besides sharing personal details and feeling less need to talk to human. LLM-D12 has only been developed and validated in English-speaking contexts, and no equivalent measure is currently available for Arabic-speaking populations. As such, no scale was adapted for Arabic-speaking populations, which means that no validated instruments are currently available to assess LLM dependency in Arabic-language contexts - in a region where adoption of generative AI is rapidly expanding.

Although another tool, the Generative AI Dependency Scale [24], have demonstrated structural invariance across two culturally distinct populations; a Western context (the United States), typically considered as high in individualism and an Eastern (Singapore), typically considered as high in collectivism, it does not account for the diversity of LLM use contexts, such as instrumental reliance versus emotional attachment. Furthermore, scalar invariance alone does not ensure that the meaning or social implications of a score are equivalent across cultures [25]. For instance, the same level of relationship dependency may carry different connotations or behavioural consequences depending on the cultural context.

Relatedly, cultural bias may arise despite flawless linguistic translation. An item such as "feeling irritable when interrupted online", an item typically used in measuring internet addiction, may reflect cultural values surrounding personal space and autonomy, which are prioritised differently across societies. Likewise, the perceived intensity of emotions or the social acceptability of certain behaviours described in items can vary significantly, leading respondents from different cultures to interpret or respond to the same question in systematically different ways [26] [27]. This is supported by research comparing cross-cultural differences between the Arab Gulf Cooperation Council (GCC) and other countries, which indicates that cultural norms in the region may shape how individuals perceive and respond to questions related to social behavior and emotional expression [28] [29] [30]. Studies also demonstrated that cultural dimensions, such as in-group collectivism and the centrality of social networks in this region, significantly predict technology adoption [31]. This suggests that LLM dependency may present differently in collectivistic societies. For instance, "social isolation," often a hallmark of addictive behaviour, may relate less to physical solitude and more to a perceived disconnection from one's primary social groups.

AI adoption in Arab countries is expanding rapidly, driven by the national strategies. In the United Arab Emirates (UAE), an Arab country in the GCC region, AI usage is reported at nearly 100% [32], and many are interacting with AI tools for emotional support and companionship companion [33]. In parallel, Arab nations have accelerated efforts to build their own Arabic-language LLMs, reflecting both regional demand and strategic technological leadership. For example, Fanar, developed in Qatar, was designed to be culturally aware and fluent in Arabic dialects [34]. Such models further expand access to culturally and linguistically relevant generative AI, marking a shift from passive adoption to active development, potentially deepening integration of LLMs into everyday digital life.

Specific cultural values in the Arab population have been also shown to influence attitudes towards work-related technology. Concepts such as 'the pursuit of excellence' and 'cooperation' have emerged as meaningful predictors of technology use [31]. These values may shape expressions of LLM dependency and they are culture-dependent: a user driven by 'the pursuit of excellence' may develop compulsive use patterns in the pursuit of perfection, while 'cooperation' may lead to dependency as a means of enhancing collaboration and sustaining group harmony. Moreover, a comparative study across 47 countries showed that emerging economies, including Arab countries like the UAE, Saudi Arabia, and Egypt, demonstrated higher trust in and use of AI compared to developed nations, which in turn was associated with less critical thinking or evaluation when using these tools [35]. Taken together, these findings suggest the importance of validating LLM dependency measures within Arab cultural contexts to ensure that the tool is both appropriate and sensitive to culturally influenced patterns of use and interpretation.

Validation of LLMs dependency assessment introduces additional challenges that extend beyond cultural sensitivity alone. First, the construct itself is novel, proposed in 2025. Unlike well-established conditions such as internet gaming disorder, there are no universally accepted "gold standard" instruments for assessing LLM dependency, making it difficult to establish criterion validity through comparison with existing measures. This also applies to the choice of external validators. Second, the broader field of internet addiction is marked by ongoing conceptual and psychometric ambiguity. Terms such as "problematic use," "compulsive use," "excessive use," and "addiction" are frequently used interchangeably, reflecting a lack of definitional consensus. This lack of definitional clarity has also prompted calls for a unified terminology, as illustrated by Almourad et al. [36], who describe Digital Addiction as a broad construct encompassing patterns of use, harm caused, and personal factors.

Given these conceptual uncertainties, alongside the distinctive cultural and technological landscape of the Arab world, there is a need for empirical work that explores how LLM dependency manifests in this context using culturally validated assessment tool. This study aims to validate an Arabic translation of the LLM-D12 [23] in an Arab sample and thoroughly evaluate its psychometric properties. Beyond the validation process, the research investigates the relationships between the two components of LLM-D12 (the Instrumental and Relationship) and several key variables, including Internet addiction, attitude toward AI, perceived trustworthiness of LLM, and need for cognition. Through these objectives, the study will provide a linguistically and culturally appropriate instrument for assessing LLM dependency in Arabic-speaking populations, enabling researchers to investigate problematic LLM use. It will also contribute to the broader understanding of LLM dependency by exploring its manifestations within a non-Western cultural context, potentially revealing unique cultural nuances or universal patterns of human-LLM interaction. Additionally, the availability of a validated Arabic LLM-D12 will facilitate cross-cultural comparisons and inform the development of culturally sensitive interventions aimed at promoting healthy and responsible LLM use in Arab communities.

## METHOD

### Data collection and sampling procedure

This study was carried out between late June to early July 2025. The participants who met the eligibility criteria were recruited using Prolific online platform (www.prolific.com). The survey was developed and distributed using SurveyMonkey ([www.surveymonkey.com](www.surveymonkey.com)). The survey was conducted anonymously, with participants providing digital informed consent before accessing the questionnaires. Participants who successfully completed the survey and did not fail attention checks received monetary compensation.

*Inclusion Criteria.* Before proceeding to the full questionnaire, individuals were asked about their engagement with the LLM they use most frequently, referred to in the study as their "Primary LLM." Participants were included if they met at least one of the following conditions: (1) they used their Primary LLM almost daily and considered themselves significantly dependent on it, (2) they used it almost daily but not relying on it significantly, or (3) they reported significant reliance despite not using it on a daily basis. Individuals who neither used their Primary LLM frequently nor reported significant reliance were excluded from the analysis. We note here that only 3.44% were excluded due to this condition. Additional eligibility requirements included being at least 18 years old and identifying as Arab in terms of culture and norms.

The majority of participants reported using ChatGPT (OpenAI), followed by Gemini, and DeepSeek. While some alternative LLMs were also mentioned, their reported usage was substantially lower. Frequency data revealed that more than half of the sample (66.80%) interacted with LLMs multiple times each day. When rating their usage frequency on a 10-point scale, most participants reported high engagement levels, with most responses clustered at the upper end (scores 8–10).

*Sample size estimation*. Sample size requirements for the Arab-sample validation were estimated based on the structure and parameter estimates from our UK CFA (12 items, 2 correlated factors, df = 39, loadings .57–.94). Using the RMSEA-based approach [37], to detect a difference between a close fit (RMSEA = .05) and a not-close fit (RMSEA = .08) with α = .05, the Arab model (df = 53), these thresholds was N = 204 (80% power). Parallel calculations indicated that with the smallest observed loading on the UK sample (.57), N = 200 is needed to estimate factor loadings with 95% CI widths ≤ ±.10, and N = 220 to estimate α = .80 with CI width ≤ .05 [38]. We therefore targeted N = 230, which exceeds the 80% RMSEA power requirement for both models, provides near-90% power for the Arab CFA, and ensures estimation of factor loadings and reliability coefficients, with enough power to detect small-to-moderate correlations.

A total of 304 responses were collected. However, after data cleaning procedure, which is described in section Data Preprocessing, the final sample included 250 participants (mean age of 28.27 years (SD = 6.08; 47.6% self-identified as male, 52.4% as female). The vast majority of participants (96.8%) were aged 18-40 or below, with a small subset (3.2%) aged between 41- and 56-year-old.

*Arabic translation of the survey*. A conceptually and linguistically valid Arabic version of the scale was developed using the back-translation method [39]. First, a bilingual author of this work, proficient in both Arabic and English, translated the original English version into Arabic. Second, another bilingual author, expert in Human-AI Collaboration and research methods, reviewed and approved the Arabic version for linguistic accuracy and fluency. Third, a different bilingual academic, not part of the authors' team and unfamiliar with the scale but experienced in survey design, translated the Arabic version back into English. The senior author then compared the back-translated English version with the original LLM-D12 to ensure that the meaning was preserved. Minor adjustments were made to reduce potential ambiguity.

A pilot study of the Arabic survey was conducted with 37 participants to test clarity and ease of understanding. Minor adjustments were made to improve sentence flow and remove unintended ambiguity. One translated item originally read: *"I use it solely as a tool, not to express my feelings or expect it to understand me."* Most Arab participants strongly agreed with this statement, probably because they focused on the first part of the sentence and overlooked the full meaning. To address this and preserve the intended reversed meaning, the item was reworded as: *"I do not expect it to understand me, nor do I share or discuss my feelings with it. I use it solely as a tool."* After this change, no further modifications were made to the translation.

**Measurements**

*Demographics.* Participants were asked about their age, gender and education.

*LLM-D12.* LLM-D12 scale consists of 12 items [23] that represent observed or theorized symptoms of dependency on LLM. The scale was developed based on the theoretical foundation proposed in [10], and it captures two theoretically distinct but related dimensions of LLM dependency:

*Instrumental Dependency* (6 items) captures the habitual reliance on LLMs for memory support, decision-making, task automation, and cognitive simplification. This type of dependency is rooted in the psychological concepts of cognitive offloading and flow states, where users delegate mental tasks to the LLM and become immersed in interaction patterns due to the LLM's responsiveness and adaptability, which could encourage continued use and task absorption.

*Relationship Dependency* (6 items) reflects users' emotional and social attachment to the LLM. Drawing on concepts of anthropomorphism and parasocial interaction, this subscale measures how users may perceive LLMs as companions, attributing human-like qualities to them and forming one-sided parasocial relationships that mimic real-world social connections. Such dependency is associated with sharing more personal details with LLM and reduces the desire or need to talk to humans.

Consistent with the original LLM-D12 design, all items in the Arabic version were rated using a 6-point Likert (1 = "Strongly Disagree" to 6 = "Strongly Agree") without a neutral midpoint. The choice of omitting a neutral option was to prompt respondents to indicate a clear position on either agreement or disagreement. The list of LLM-D12 items in both English and Arabic is provided in Appendix 1.

*Internet Addiction (IA)*. IA was assessed using the 7-item short form of the Internet Addiction Test (IAT-7)[40]. Participants rated each item on a 5-point Likert scale ranging from 1 ("Never") to 5 ("Always"), with a total score range between 7 and 35. The Cronbach's alpha for this scale in our sample was 0.82.

*Attitudes Toward AI (ATAI)*. ATAI includes 2 subscales: ATAI Acceptance (2 items) and ATAI Fear (3 items) [41]. All items were rated on an 11-point Likert scale from 0 ("Strongly Disagree") to 10 ("Strongly Agree"). The total score for ATAI Acceptance ranged from 0 to 20, and for ATAI Fear from 0 to 30. Cronbach's alpha in the present study was 0.69 and 0.74, respectively.

*Need for Cognition (NFC)*. NFC was measured using the 6-item short version of the NFC scale [59]. Responses were rated on a 5-point Likert scale ranging from 1 ("Extremely uncharacteristic of me") to 5 ("Extremely characteristic of me"), with total scores ranging from 6 to 30. The internal consistency of this scale in our study was 0.79.

*Perceived Trustworthiness of primary LLM*. Perceived Trustworthiness of primary LLM was evaluated using an 8-item scale adapted from a prior established framework [43], covering eight dimensions: truthfulness, safety, fairness, robustness, privacy, machine ethics, transparency, and accountability. Each item was assessed on an 11-point Likert scale (0 = "Not at all" to 10 = "Completely"). An example item is: "I trust my primary LLM to provide factually accurate and reliable information." Scores were summed to obtain a total trust score ranging from 0 to 80. The Cronbach Alpha for this measure in the present study was 0.88.

*The use of LLMs*. Participants were also asked to report their primary LLM and their frequency of use of it, rated on an 11-point scale from 0 ("Very infrequently") to 10 ("Very frequently"). Additionally, they reported the method of interaction with their primary LLM with response options included: text, voice, text and voice, and other. Participants also reported on the device they primarily use to access their LLM, with options: PC, mobile, tablet, smart speaker (e.g., Amazon Echo, Google Nest), dedicated AI device (e.g., Humane AI Pin, Rabbit R1), and other. Finally, participants indicated the typical location of LLM use, choosing from: stationary (e.g., at home, office, school), on the move (e.g., while commuting, traveling), both - stationary and on the move, and other.

*Attention check*. The survey included attention check items to ensure data quality. These items were carefully framed to match the general tone, structure and length of other scale items, making them less detectable and helping to identify inattentive respondents.

**Data analysis**

*Data preprocessing*. The original sample comprised 304 participants who fully completed the survey. All questions were mandatory, so completing the survey means no answers were left missing. We removed participants who failed attention checks (n = 30, 9.9%). The data were then examined for duplicate entries, resulting in the exclusion of 24 cases (7.9%). Duplicates were identified using the Prolific ID. The final sample for analysis included 250 participants.

*Monotonicity analysis*. None of the LLM-D12 scale items contained missing values. To evaluate the scalability and appropriateness of our proposed scales, we conducted nonparametric Mokken scale analysis via the mokken package (version 3.1.2 [44]) in R. Mokken scaling is a probabilistic approach within item response theory that assesses whether the probability of endorsing an item increases monotonically as the latent trait increases which is a key requirement for ordinal scales. For the Instrumental

Dependency scale, the number of observed violations was zero. The item scalability coefficients (Hi) for this scale ranged from 0.42 to 0.58, with all coefficients exceeding the commonly accepted threshold of 0.3, thereby supporting the internal scalability of the scale. Similarly, the Relationship Dependency scale showed no violations of the monotonicity assumption for any item. Scalability coefficients (Hi) for the Relationship Dependency items ranged from 0.47 to 0.71, exceeding the accepted cut-off.

*Confirmatory Factor Analysis (CFA)*. CFA was used to test the hypothesized two-factor structure comprising the Instrumental Dependency and Relationship Dependency. Model fit was evaluated using multiple indices, including the Comparative Fit Index (CFI), Tucker-Lewis Index (TLI), Root Mean Square Error of Approximation (RMSEA) with confidence interval, Standardized Root Mean Square Residual (SRMR), and the chi-square statistic. Potential multivariate outliers were tested with Mahalanobis distance. Standardized factor loadings and item-level explained variance ($R^2$) were calculated to evaluate the strength of association between observed items and their respective latent constructs.

*Internal consistency*. Internal consistency for each scale was evaluated using Cronbach's alpha. Item-level reliability was further examined by calculating corrected item-total correlations, which assess the degree to which each item relates to the sum of the remaining items within its scale.

*Discriminant validity*. Discriminant validity between the Instrumental Dependency and Relationship Dependency constructs was examined using multiple complementary approaches. First, the Fornell-Larcker criterion was applied, whereby the average variance extracted (AVE) for each construct is compared with the squared correlation between the constructs; discriminant validity is supported when the AVE values exceed the squared correlation. Second, the Heterotrait–Monotrait Ratio of correlations (HTMT) was calculated. HTMT assesses discriminant validity by comparing the mean of correlations between items measuring different constructs with the mean of correlations between items measuring the same construct. Values below the commonly recommended threshold of 0.85 are considered evidence that constructs are empirically distinct. Finally, a chi-square difference test was conducted to compare the fit of the hypothesised two-factor model with an alternative one-factor model in which all items were loaded onto a single latent construct. Differences in model fit were evaluated using $\chi^2$ difference statistics, Akaike Information Criterion (AIC), Bayesian Information Criterion (BIC), and RMSEA, with lower AIC/BIC values and smaller RMSEA values indicating better fit.

*Network analysis*. A network analysis was carried out to examine the structural validity of the Arabic version of the scale, using all 12 items from the Instrumental and Relationship Dependency subscales. The network structure was estimated with the EBICglasso algorithm, which regularises partial correlations to produce a sparse and interpretable network. Centrality measures were then calculated to indicate the relative contribution of each item within the network. Analyses were conducted in R (v4.3.1) using bootnet (function *estimateNetwork*); correlations for ordinal items were computed via *qgraph*'s cor_auto. Network visualisation and centrality indices were obtained with qgraph.

*External validation*. To evaluate the convergent and discriminant validity of the Instrumental Dependency and Relationship Dependency subscales, we examined associations between each dependency subscale and a set of theoretically relevant variables, including AI acceptance, AI fear, internet addiction, need for cognition, and perceived trustworthiness of LLM using Multivariate Multiple Regression. All variables were standardized prior to analysis to ensure comparability across different measurement scales. Post hoc Wald tests were used to compare the strength of each predictor's association between the two dependency outcomes. A post hoc sensitivity analysis was conducted to evaluate whether the sample size was adequate to detect the observed effect sizes from the multivariate multiple regression models used in the external validation stage. Effect sizes were calculated as Cohen's $f^2$ from the observed $R^2$ values for each model, with k representing the number of predictors. Using these f² values, statistical power was estimated with α set at 0.05 (two-tailed) to determine the probability of detecting the overall model effects given the achieved sample size.

*Additional analysis*. We tested whether patterns of LLM use were uniquely associated with each form of dependency. LLM use variables included frequency of use, primary location of use, type of device, and preferred interaction method. All associations were examined using multivariate multiple regression, with the two dependency scores entered as simultaneous outcome variables.

**RESULTS**

**Descriptive statistics**

Means and standard deviations for each item of the Instrumental Dependency and Relationship Dependency scales are presented in Table 1.

**Table 1**. Descriptive Statistics for the Instrumental and Relationship Dependency Scale Items

| Instrumental Dependency | | | Relationship Dependency | | |
|---|---|---|---|---|---|
| *Item Code* | *Mean* | *SD* | *Item Code* | *Mean* | *SD* |
| Without_it_less_confident | 2.988 | 1.343 | Share_private_life | 3.160 | 1.625 |
| Immersed_in_it | 3.436 | 1.491 | Genuine_companion | 2.880 | 1.596 |
| More_rewarded | 4.032 | 1.317 | Feel_less_alone | 2.640 | 1.637 |
| Turn_to_it_even_when_able | 3.692 | 1.410 | It_adds_to_my_social_life | 2.464 | 1.524 |
| It_is_my_go-to | 3.764 | 1.447 | Solely_as_a_tool | 2.868 | 1.634 |
| Decisions_without_uneasy | 2.676 | 1.384 | No_need_to_talk_to_others | 2.404 | 1.467 |

To estimate the precision of item means, 95% confidence intervals were calculated for each item, providing a range that is likely to contain the true mean score for the population (Figure 1). The plot indicates that mean scores for the Instrumental Dependency items were generally higher than those for the Relationship Dependency items. All confidence intervals were narrow, reflecting a high degree of agreement among respondents for each item.

**Figure 1**. Item means with 95% confidence intervals for the instrumental and relationship dependency scales

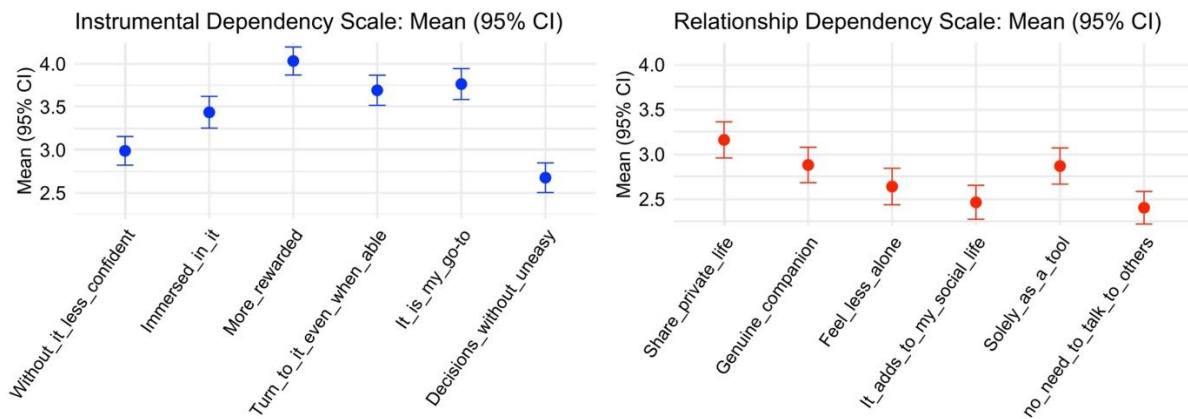

**Confirmatory Factor Analysis**

Confirmatory factor analysis was performed to evaluate the measurement model consisting of the Instrumental Dependency and Relationship Dependency scales. The model demonstrated good fit CFI = 0.95, TLI = 0.93 (exceeding the commonly recommended threshold of 0.90) [45]. The RMSEA was 0.08, with a 90% confidence interval of [0.07, 0.10]. The SRMR was 0.06, indicating a good fit (values ≤ 0.08 are generally considered acceptable). A chi-square statistic of 145.30 (df = 53, p < .001), indicating a significant difference from a perfectly fitting model; however, the chi-square test is known to be sensitive to sample size and minor deviations from model assumptions.

Prior to model estimation, multivariate outliers were identified using Mahalanobis distance. Six participants (2.4% of the sample) were flagged as potential outliers; however, analyses conducted with and without these cases showed substantively similar model fit statistics. Therefore, all results reported below include the full sample.

All items demonstrated strong and significant standardized factor loadings (all p < .001) on their respective latent constructs (Table 2). All loadings exceeded the conventional threshold of 0.50. For the Instrumental Dependency factor, loadings ranged from 0.53 to 0.82, for the Relationship Dependency factor, standardized loadings ranged from 0.54 to 0.92. These results suggest that each item serves as a reliable indicator of its latent factor.

The proportion of explained variance ($R^2$) for each item was examined to assess the degree to which each observed variable reflected its underlying latent factor. For the Instrumental Dependency items, $R^2$ values ranged from 0.28 to 0.67, indicating that between 28% and 67% of the variance in individual items was explained by the Instrumental Dependency factor. Uniqueness values (1 - $R^2$) for these items ranged from 0.33 to 0.72, suggesting moderate amounts of residual item-specific variance.

For the Relationship Dependency items, $R^2$ values were generally higher, ranging from 0.29 to 0.85. This suggests that the Relationship Dependency factor accounted for between 29% and 85% of the variance in its respective indicators, with uniqueness values ranging from 0.15 to 0.71. Notably, the items "Feel_less_alone," "It_adds_to_my_social_life," and "No_need_to_talk_to_others" demonstrated particularly high proportions of explained variance ($R^2 > 0.74$).

Taken together, results suggest that all of scale items are well-represented by their latent constructs. Items with lower $R^2$ values (e.g., "More_rewarded" and "Solely_as_a_tool") may reflect unique variance not fully captured by the factors, but they remain within acceptable limits for inclusion in the measurement model.

**Table 2**. Standardized factor loadings, standard errors (SE), uniqueness, and explained variance ($R^2$) for items of the Instrumental and Relationship Dependency

| Scale | Item code | Loading | SE | z (p<.001) | Uniqueness | $R^2$ |
|---|---|---|---|---|---|---|
| Instrumental Dependency | Without_it_less_confident | 0.748 | 0.033 | 22.627 | 0.441 | 0.559 |
| | Immersed_in_it | 0.618 | 0.044 | 14.166 | 0.618 | 0.382 |
| | More_rewarded | 0.530 | 0.050 | 10.647 | 0.719 | 0.281 |
| | Turn_to_it_even_when_able | 0.816 | 0.027 | 29.919 | 0.335 | 0.665 |
| | It_is_my_go_to | 0.812 | 0.028 | 29.455 | 0.341 | 0.659 |
| | Decisions_without_uneasy | 0.681 | 0.039 | 17.625 | 0.536 | 0.464 |
| Relationship Dependency | Share_private_life | 0.652 | 0.038 | 16.993 | 0.575 | 0.425 |
| | Genuine_companion | 0.771 | 0.028 | 27.684 | 0.406 | 0.594 |
| | Feel_less_alone | 0.924 | 0.013 | 72.423 | 0.145 | 0.855 |
| | It_adds_to_my_social_life | 0.906 | 0.014 | 62.492 | 0.180 | 0.820 |
| | Solely_as_a_tool | 0.538 | 0.047 | 11.453 | 0.711 | 0.289 |
| | No_need_to_talk_to_others | 0.862 | 0.019 | 45.755 | 0.258 | 0.742 |

Nore: the z-score represents the standardized estimate of each item's loading divided by its standard error, indicating how strongly each item is associated with its factor relative to the variability of that estimate. P-value for all items were <.001.

## Validation

### Internal consistency

Internal consistency was assessed for each scale using Cronbach's alpha. The Instrumental Dependency scale demonstrated good reliability with Cronbach's alpha of 0.85 across its six items, while the Relationship Dependency scale showed excellent reliability (alpha = 0.90). When considering all twelve items together, the overall scale showed excellent internal consistency with Cronbach's alpha of 0.90. These findings support the adequacy of each subscale and the total scale for research and applied use, indicating that the items within each scale are measuring coherent, underlying constructs.

Corrected item-total correlations for the Instrumental Dependency scale ranged from 0.52 to 0.72. All items showed correlations above the conventional acceptability threshold of 0.30, suggesting that each item was positively and substantially related to the overall scale score (i.e., the sum of the remaining items). Notably, "More_rewarded" had the lowest item-total correlation (0.52), but this value remains within acceptable limits, suggesting the item captures relevant aspects of Instrumental Dependency. Corrected item-total correlations for the Relationship Dependency scale ranged from 0.50 to 0.81, demonstrating that each item contributes positively to the measurement of the underlying construct and correlates well with the overall scale (Table 3).

Taken together, this analysis suggests good internal consistency and reliability of both scales.

**Table 3**. Internal consistency and item analysis indices for Instrumental Dependency and Relationship Dependency

| Scale | Item code | Chronbach's alpha | alpha SE | r.cor | var.r | r.drop |
|---|---|---|---|---|---|---|
| Instrumental Dependency | Without_it_less_confident | 0.896 | 0.010 | 0.689 | 0.026 | 0.661 |
| | Immersed_in_it | 0.898 | 0.010 | 0.651 | 0.028 | 0.598 |
| | More_rewarded | 0.904 | 0.009 | 0.522 | 0.023 | 0.458 |
| | Turn_to_it_even_when_able | 0.894 | 0.010 | 0.720 | 0.026 | 0.698 |
| | It_is_my_go_to | 0.896 | 0.010 | 0.696 | 0.025 | 0.674 |
| | Decisions_without_uneasy | 0.896 | 0.010 | 0.687 | 0.027 | 0.653 |
| Relationship Dependency | Share_private_life | 0.895 | 0.010 | 0.691 | 0.026 | 0.656 |
| | Genuine_companion | 0.890 | 0.010 | 0.770 | 0.024 | 0.754 |
| | Feel_less_alone | 0.887 | 0.011 | 0.804 | 0.021 | 0.811 |

| | | | | | |
|---|---|---|---|---|---|
| It_adds_to_my_social_life | 0.887 | 0.011 | 0.810 | 0.022 | 0.815 |
| Solely_as_a_tool | 0.906 | 0.009 | 0.508 | 0.020 | 0.443 |
| no_need_to_talk_to_others | 0.889 | 0.010 | 0.794 | 0.023 | 0.792 |

*Note*: alpha SE represents the standard error of the Cronbach's alpha estimate. r.cor - the item-total correlation after correcting for item overlap (i.e., the correlation between the item and the total of the remaining items, not including itself). var.r - the variance of the item-item correlations with low values suggest item homogeneity. r.drop - corrected item-total correlation if the item dropped

*Discriminant validity*

Discriminant validity between the Instrumental Dependency and Relationship Dependency scales was assessed using several complementary approaches. The Fornell-Larcker criterion indicated that the average variance extracted (AVE) was 0.505 for the Instrumental Dependency construct and 0.613 for the Relationship Dependency construct. Both values exceeded the squared correlation between the latent factors (0.37), supporting discriminant validity. The HTMT value was 0.64, well below the recommended threshold of 0.85, indicating that the constructs are empirically distinct.

A chi-square difference test was conducted to compare the fit of the hypothesized two-factor model to the one-factor model in which all items loaded onto a single latent factor. The two-factor model provided a significantly better fit, $\Delta\chi^2(1) = 313$, $p < .001$. The two-factor solution also showed superior fit indices (AIC = 9258.9, BIC = 9347.0, RMSEA = 0.08) compared to the one-factor model (AIC = 9570.0, BIC = 9654.5, RMSEA = 1.12), further supporting the distinction between the two constructs.

*Network analysis*

To estimate the structure of both scales, we performed network analysis. Network centrality analysis for the Instrumental Dependency items showed that "It_is_my_go_to" (1.12) and "Turn_to_it_even_when_able" (1.08) were the most central nodes in the network. These were followed by "Without_it_less_confident" (1.01), "Decisions_without_uneasy" (0.97), "More_rewarded" (0.72), and "Immersed_in_it" (0.67). Closeness centrality values that show how quickly a particular item (node) can reach all other items in the network via the shortest possible paths, were highest for "It_is_my_go_to" (0.0082) and "Turn_to_it_even_when_able" (0.0079), indicating that these two items are not only highly connected, but also efficiently linked to the rest of the network. Taken together, the results suggest that feeling that LLM is one's "go-to" option and frequently turning to it, are the most structurally important features of Instrumental Dependency in this sample (Figure 2).

Analysis of network centrality for the Relationship Dependency scale indicated that "Feel_less_alone" (1.26) and "It_adds_to_my_social_life" (1.24) emerged as hubs within this scale. "No_need_to_talk_to_others" (0.95) and "Genuine_companion" (0.93) also showed substantial levels of connectivity, suggesting their relevance in capturing interpersonal aspects of Relationship Dependency (Figure 2). In terms of closeness centrality, "Feel_less_alone" (0.0087), "It_adds_to_my_social_life" (0.0086), and "No_need_to_talk_to_others" (0.0086) had the highest values among the relationship items, showing that these nodes are positioned to most efficiently connect with the rest of the network. In contrast, "Share_private_life" (degree = 0.68, closeness = 0.0057) and "Solely_as_a_tool" (degree = 0.60, closeness = 0.0067) were less connected and less centrally positioned. Taken together, these results suggest that perceptions of reduced loneliness, the belief that LLM adds to one's social life, and diminished need to talk to others are the most structurally influential and efficiently interconnected aspects of Relationship Dependency in this sample.

**Figure 2**. Network structure of Instrumental and Relationship Dependency items

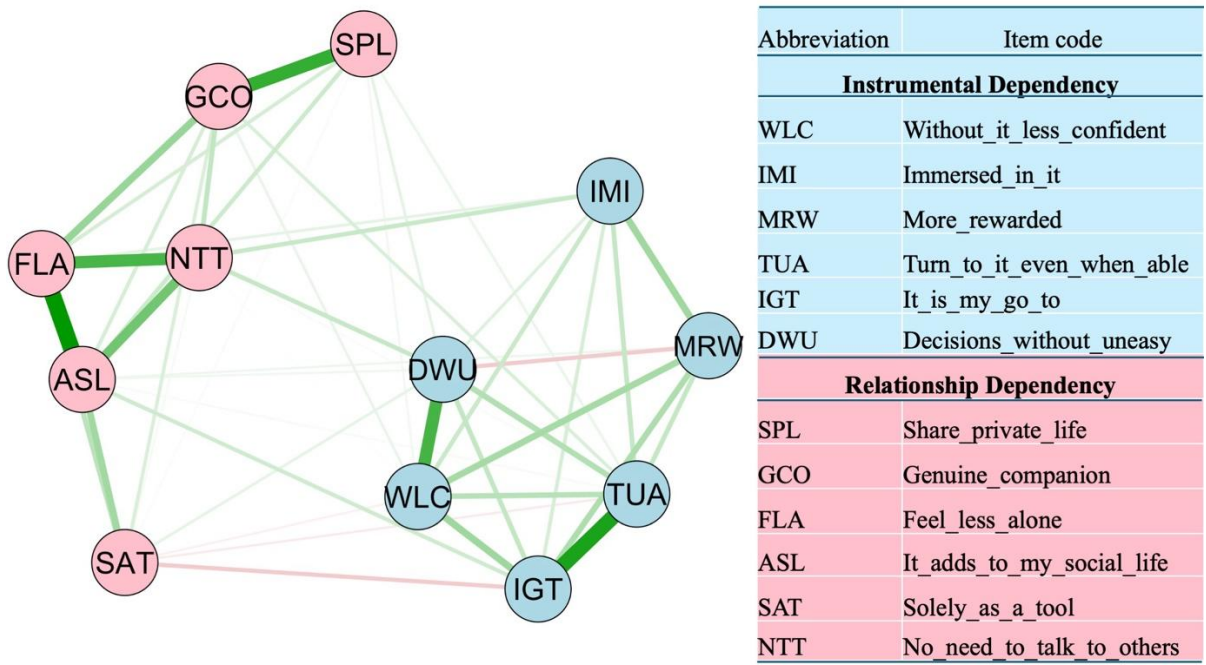

| Abbreviation | Item code |
|---|---|
| **Instrumental Dependency** | |
| WLC | Without_it_less_confident |
| IMI | Immersed_in_it |
| MRW | More_rewarded |
| TUA | Turn_to_it_even_when_able |
| IGT | It_is_my_go_to |
| DWU | Decisions_without_uneasy |
| **Relationship Dependency** | |
| SPL | Share_private_life |
| GCO | Genuine_companion |
| FLA | Feel_less_alone |
| ASL | It_adds_to_my_social_life |
| SAT | Solely_as_a_tool |
| NTT | No_need_to_talk_to_others |

*External validation*

    Multivariate multiple regression analysis was performed to examine the simultaneous associations between AI Acceptance (ATAI Acceptance), AI fear (ATAI Fear), Internet Addiction, Need for Cognition, and Perceived Trustworthiness of LLM and both Instrumental Dependency and Relationship Dependency scores. The overall model was significant, Wilks' $\Lambda = 0.770$, $F(10, 486) = 6.77$, $p < .001$, indicating that the combined predictors explained a significant amount of variance across the two outcomes.

    Follow-up univariate regression analyses were then examined for each outcome separately. For Instrumental Dependency, the model accounted for 19.7% of the variance ($R^2 = .197$, Adjusted $R^2 = .180$), $F(5, 244) = 11.95$, $p < .001$. Higher AI acceptance ($\beta = 0.23$, SE = 0.08, $p = .003$) and higher Internet Addiction ($\beta = 0.27$, SE = 0.06, $p < .001$) were significant positive predictors. For Relationship Dependency, the model explained 14.4% of the variance ($R^2 = .144$, Adjusted $R^2 = .126$), $F(5, 244) = 8.19$, $p < .001$. Significant predictors included Internet Addiction ($\beta = 0.17$, SE = 0.06, $p = .007$), Perceived Trustworthiness of LLM ($\beta = 0.19$, SE = 0.08, $p = .018$), and lower need for cognition ($\beta = -0.13$, SE = 0.06, $p = .035$) (Table 4).

**Table 4**. Results of Multivariate Multiple Regression

| Outcome | Predictor | Estimate | SE | t-test | p-value |
|---|---|---|---|---|---|
| **Instrumental Dependency** | | | | | |
| | *AI Acceptance* | *0.230* | *0.077* | *3.000* | *0.003* |
| | AI Fear | -0.090 | 0.062 | -1.453 | 0.148 |

|  | | | | |
|---|---|---|---|---|
| NFC | -0.073 | 0.061 | -1.212 | 0.227 |
| ***Internet Addiction*** | ***0.271*** | ***0.060*** | ***4.514*** | ***0.000*** |
| Perceived Trustworthiness of LLM | 0.132 | 0.075 | 1.746 | 0.082 |
| **Relationship Dependency** | | | | |
| AI Acceptance | 0.142 | 0.079 | 1.800 | 0.073 |
| AI Fear | -0.119 | 0.064 | -1.871 | 0.062 |
| ***NFC*** | ***-0.132*** | ***0.063*** | ***-2.116*** | ***0.035*** |
| ***Internet Addiction*** | ***0.167*** | ***0.062*** | ***2.703*** | ***0.007*** |
| ***Perceived Trustworthiness of LLM*** | ***0.186*** | ***0.078*** | ***2.387*** | ***0.018*** |

*Note*: Estimate represents the standardized regression coefficient (β), indicating the direction and strength of the association between the predictor and the outcome, controlling for all other predictors. SE - the standard error of the estimate

In addition, we performed Post Hoc Wald z-tests to examine whether each predictor had a differential effect on Instrumental Dependency and Relationship Dependency. The results showed that none of the predictors showed a statistically significant difference in their effects, suggesting that all predictors operate similarly in influencing both scales (Table 5).

**Table 5.** Wald test comparing predictor effects across Instrumental and Relationship Dependency

| Predictor | Instrumental | Relationship | Diff | SE Diff | z | p | CI l | CI up |
|---|---|---|---|---|---|---|---|---|
| AI Acceptance | 0.230 | 0.142 | 0.087 | 0.110 | 0.794 | 0.427 | -0.128 | 0.303 |
| AI Fear | -0.090 | -0.119 | 0.030 | 0.089 | 0.334 | 0.739 | -0.145 | 0.204 |
| Internet Addiction | 0.271 | 0.167 | 0.103 | 0.086 | 1.199 | 0.230 | -0.066 | 0.272 |
| NFC | -0.073 | -0.132 | 0.059 | 0.087 | 0.677 | 0.499 | -0.112 | 0.230 |
| Perceived Trustworthiness of LLM | 0.132 | 0.186 | -0.054 | 0.108 | -0.500 | 0.617 | -0.267 | 0.158 |

*Note*: Diff represents the difference between the Instrumental and Relationship coefficients; SE Diff is the standard error of the difference; z is the Wald test statistic for the difference; p is the p-value testing whether the difference is significantly different from zero; CI l and CI up represent the lower and upper bounds of the 95% confidence interval for the difference

After performing external validation of our scales, we ran a sensitivity analysis to estimate the adequacy of the Arab-sample size based on the observed effect sizes from the multivariate multiple regression models. For the Instrumental Dependency model ($R^2 = 0.197$, k = 5 predictors, N = 250), the corresponding Cohen's $f^2$ was 0.245, which falls between Cohen's [46] thresholds for a medium effect (0.15) and a large effect (0.35). For the Relationship Dependency model ($R^2 = 0.144$), $f^2$ was 0.168, indicating a medium effect. Using these values, and with α = 0.05 (two-tailed), the achieved statistical power was >0.95 and ≈0.98, respectively, for detecting the observed overall model effects [47]. These results confirm that the achieved sample size was more than sufficient to detect medium-to-large multivariate effects in the external validation stage.

### Additional analysis

To examine whether patterns of LLM usage were associated with levels of Instrumental and Relationship Dependency, we carried out a multivariate multiple regression analysis. Predictor variables included frequency of LLM use, primary location of use (stationary versus stationary and on the move), type of device (PC-based versus mobile devices), and interaction method (text only versus text and voice) The device, location, and interaction method were analysed as categorical variables. Prior to analysis, categories with insufficient responses were excluded to ensure reliable estimation.

The overall model revealed significant multivariate effects for frequency of LLM use (Wilks' $\Lambda$ = 0.95, $F(2, 221)$ = 5.69, $p$ = .004) and interaction method (Wilks' $\Lambda$ = 0.88, $F(2, 221)$ = 14.86, $p < .001$), indicating that these predictors were associated with dependency outcomes. In contrast, device type (Wilks' $\Lambda$ = 0.98, $F(2, 221)$ = 2.17, $p$ = .12) and location of use (Wilks' $\Lambda$ = 0.99, $F(2, 221)$ = 0.503, $p$ = .97) were not significant multivariate predictors. These results indicate that, in this sample, frequency and method of LLM interaction are the primary contextual variables linked to dependency scores (Table 6).

**Table 6**. Mean and standard deviation (in brackets) for Instrumental and Relationship Dependency across LLM usage

|  | Device | | Interaction method | | Location | | Frequency of LLM use |
|---|---|---|---|---|---|---|---|
|  | Mobile | PC-based | Text | Text & Voice | Stationary | Stationary and on the move |  |
| Instrumental Dependency | 20.65 (6.91) | 20.54 (5.91) | 19.96 (6.09) | 22.00 (6.74) | 20.02 (6.34) | 21.1 (6.34) | 20.58 (6.35) |
| Relationship Dependency | 17.23 (8.44) | 15.78 (7.10) | 14.98 (7.36) | 19.65 (7.63) | 15.28 (7.45) | 17.43 (7.87) | 16.41 (7.73) |

Post Hoc test showed that for the Instrumental Dependency, frequency of LLM use emerged as the significant predictor ($b$ = 0.83, SE = 0.22, $t$=3.62, $p < .001$), indicating that more frequent use of large language models was associated with higher Instrumental Dependency scores. Interaction method was also significant ($b$=3.57, SE = 0.97, $t$=3.68, $p<.001$) showing that individuals who use both Text and Voice for interaction with LLMs have higher scores on the Instrumental Dependency scale. Device type and location of use has no significant associations with Instrumental Dependency (all $p > .05$). The overall model accounted for 9.1% of the variance in Instrumental Dependency scores (adjusted $R^2$ = 0.091, $F(4, 222)$ = 6.67, $p < .001$).

For the Relationship Dependency, the overall model explained approximately 12.3% of the variance in Relationship Dependency (adjusted $R^2$ = 0.123, $F(4, 222)$ = 8.89, $p < .001$). In contrast to Instrumental Dependency, Post Hoc test revealed that only interaction method was a significant predictor of the relationship Dependency score ($b$=6.06, SE = 1.18, $t$=5.14, $p<.001$). All other predictors (device type, frequency of LLM use, location of use) did not reach significance (all $p > .05$).

### DISCUSSION

LLM dependency is a new and evolving phenomenon, with individuals increasingly relying on these systems not only for cognitive support but also for companionship. As this form of reliance becomes more embedded in daily life, there is a growing need for validated tools that can reliably capture its distinct psychological dimensions and be culturally sensitive. This study addresses that gap by validating a dual-dimensional scale designed to assess Instrumental and Relationship Dependency on LLM in an Arab sample.

The original LLM-D12 scale was developed and validated within a UK population [23]. For the present validation study, we employed an Arab sample with comparable demographic characteristics, including a balanced proportion of males and females and a similar age distribution.

**Psychometric evaluation of the Arabic version of the LLM-D12**

The Arabic translation of the LLM-D12 demonstrated strong psychometric properties. Monotonicity checks revealed no violations among the retained items, indicating that participants consistently used the response options in a manner that corresponded to increasing levels of dependency. This suggests that the items were interpreted as intended and performed reliably within the new linguistic and cultural context. Confirmatory factor analysis further supported the structural validity of the scale, replicating the original two-factor structure representing instrumental and relationship dependency. In addition, a test for common method bias indicated that a single-factor model accounted for significantly less variance than the two-factor solution, suggesting that common method variance is unlikely to have compromised the validity of the results. Both subscales also demonstrated very good internal consistency, with Cronbach's alpha coefficients exceeding the commonly accepted threshold for reliability, indicating that the items within each subscale measured coherent and internally stable constructs.

Overall, these findings indicate that the Arabic version of the LLM-D12 is a psychometrically sound and culturally appropriate measure of instrumental and relationship dependency, suitable for use in Arab populations with demographic profiles comparable to the original UK sample.

**Instrumental and Relationship Dependency scales in Arab sample**

All six items of the Instrumental Dependency subscale demonstrated factor loadings in the Arab sample comparable to those observed in the original UK validation [23]. The strongest loadings were recorded for the items *"Turn to it even when able"* and *"It is my go-to"*, indicating consistent endorsement of core instrumental dependency behaviours across samples. Minor differences were observed, including a slightly lower loading for *"Immersed in it"* in the Arab sample (0.618 vs. 0.715), and a marginally higher loading for *"Less confident without it"* (0.748 vs. 0.706). As in the original study, the item reflecting a sense of reward from LLM use showed a relatively lower factor loading compared to other Instrumental items, though it retained strong item-total and inter-item correlations. This suggests that the rewarding aspect of LLM engagement contributes meaningfully to instrumental dependency in both cultural contexts, despite being somewhat less central to the underlying construct.

Relationship Dependency items also yielded loadings that were largely consistent with the original scale. However, the item assessing whether the LLM is used *solely as a tool* showed a notably lower loading in the Arab sample (0.538) compared to the UK sample (0.646). Two possible explanations may account for this reduction. First, participants in the Arab sample may be more likely to engage with LLMs in ways that serve social or relationship functions, consistent with the collectivist orientation of Arab cultures. Previous research has highlighted the role of in-group collectivism and social connectedness in technology adoption across Arab populations [48]. Supporting this, the item *"Add to social life"* showed a substantially higher loading in the Arab sample (0.906) than in the original scale (0.772), indicating greater emphasis on the LLM's perceived social utility. Second, analysis of the network structure revealed strong associations between *"Add to social life"* and the Instrumental item *"It is my go-to"*, which may have attenuated the unique variance accounted for by the *"Solely as a tool"* item. It is also possible that this item was interpreted differently in Arabic, despite accurate translation. The term "tool" may carry more impersonal or restrictive connotations in Arabic, potentially clashing with how LLMs are experienced when used in more exploratory or conversational ways.

Taken together, these results indicate that the items of both subscales functioned as intended in the Arab sample, supporting the cross-cultural applicability of the LLM-D12 scale.

**The relationship between Instrumental Dependency and Relationship**

Conceptually, Instrumental Dependency reflects users' reliance on LLMs for cognitive and task-related purposes, such as problem-solving, decision-making, and information retrieval. In contrast, Relationship Dependency captures a more affective, socially oriented form of engagement, wherein users interact with LLMs in ways that mimic or substitute aspects of human connection. Our results suggest that the scales were moderately correlated, consistent with the pattern observed in the original UK validation. Moreover, the relationship between them is consistent with theoretical models in human-computer interaction that acknowledge the interplay between functional and relational uses of technology ([49]; [50]). However, our findings showed that they are sufficiently distinct to warrant separate consideration in both research and applied contexts. In particular, discriminant validity analyses indicated that the Average Variance Extracted for each subscale exceeded the squared correlation between them, suggesting that each construct captures more unique variance than they share. This was further supported by a chi-square difference test, which showed a significant deterioration in model fit when the two dimensions were collapsed into a single latent factor, providing further support for two distinct factors. Complementary evidence from network analysis also demonstrated a clear structural separation between items comprising the Instrumental and Relationship Dependency scales.

The distinction between instrumental and relationship forms of dependency may offer useful avenues for both research and applied use of the LLM-D12. Researchers interested in exploring specific facets of users' engagement with LLMs might consider using the two subscales independently, depending on their theoretical or empirical focus. For example, studies targeting the cognitive or task-related dimensions of AI use may benefit from concentrating on the Instrumental Dependency scale, whereas those investigating affective, interpersonal, or socially compensatory aspects may find the Relationship Dependency scale more relevant. In applied contexts, practitioners may find it useful to calculate total scores for each subscale and interpret them as continuous measures of dependency. Where a more nuanced understanding is required, item-level responses, rated on a 6-point Likert scale, could be examined to capture finer-grained patterns within each scale. This may support more targeted assessments, particularly in settings where the nature and extent of an individual's reliance on LLMs is of practical concern, such as in educational, clinical, or workplace environments.

In addition, this distinction can support the design of interventions, policies, or tools tailored to specific forms of LLM reliance. For instance, in educational or workplace settings, elevated instrumental dependency might signal support needs or productivity enhancement, whereas high relationship dependency, might reflect a range of motivations, from compensating for limited social interaction to support in enhancing social interaction. Drawing a parallel from social media use, recent findings suggest that social media use can contribute to both problematic use and wellbeing outcomes [51][52]. This complexity highlights the importance of further to clarify how these forms of dependency relate to actual behaviours, including patterns of use, functional outcomes, and potential maladaptive reliance. At the item level, aspects such as immersion, reward-seeking, and parasocial bonding could be targeted through preventive strategies. One such approach is psychological inoculation, in which individuals are exposed to simulated scenarios to build resistance against overreliance on such patterns, as shown in recent work on building resistance against problematic social media use [53].

**Psychological and contextual correlates of LLM Dependency in the Arab sample**

Our results showed that higher Instrumental and Relationship Dependency scores were associated with greater Internet Addiction, suggesting that individuals who report problematic internet use are more likely to rely on LLMs for both tasks and companionship. However, these scales showed differential associations with AI acceptance, Perceived Trustworthiness of LLM and Need for Cognition. Specifically, individuals who showed higher AI acceptance tend to demonstrate higher Instrumental Dependency. These findings broadly align with the theoretical expectations underpinning the LLM-D12 scale [54], while also offering points of contrast with previously published data from a UK sample. In the UK, AI Acceptance was reported to be a strong and consistent predictor of both Instrumental and Relationship Dependency, with particularly pronounced effects on the latter [23]. In the Arab sample, however, AI Acceptance was associated only with Instrumental Dependency. This divergence may reflect cultural variation in the way positive attitudes towards AI are operationalised. For instance, in contexts where LLMs are more routinely embedded within everyday digital practices, such as the UK, acceptance of AI may translate more readily into both pragmatic and interpersonal forms of engagement. By contrast, in the Arab context, acceptance appears to facilitate functional use [55] [56], whereas the formation of more relationship dependencies may be shaped by other factors.

Perceived Trustworthiness of LLM, was considerably more influential in our sample than in the UK [23]. In cultural contexts characterised by collectivist values and heightened sensitivity to relational dynamics and interpersonal trust [57], trust may serve a key role in shaping users' perceptions of LLMs not merely as functional tools, but as quasi-social partners [58]. This finding suggests that affective forms of engagement with LLMs in the Arab context are more contingent upon perceived trustworthiness and reliability. In addition, Need for Cognition was negatively associated with Relationship Dependency, suggesting that individuals with lower motivation for effortful thinking may be more inclined to engage with LLMs in socially compensatory or affective ways. Despite the differences in individual predictors, our analysis revealed no significant variation in their effects across the two subscales. This may indicate that psychological antecedents of Instrumental and Relationship Dependency are less distinctly separated in the Arab context. One possible explanation is that the boundary between functional and relational uses of LLMs is more fluid, potentially reflecting broader cultural tendencies to engage with technology in socially meaningful ways, even when the context is primarily instrumental.

Another finding of potential interest to practitioners is the differential association between each scale and characteristics of LLM usage. More frequent use of LLMs was linked to higher scores on the Instrumental Dependency subscale, indicating greater reliance on these systems for practical or task-oriented purposes. Furthermore, individuals who interacted with LLMs using both text and voice modalities, as opposed to text alone, also reported higher Instrumental Dependency scores, suggesting that multimodal interaction may enhance the perceived utility and accessibility of LLMs, which potentially contribute to higher dependency. In contrast, device type and location of use were not significantly associated with Instrumental Dependency. For Relationship Dependency, interaction method was the only significant predictor, with those using both text and voice input reporting higher scores than users engaging through text alone. This indicates that relationship forms of engagement with LLMs in the Arab sample may depend more on the nature of interaction than on frequency or physical context of use. Notably, in the original scales validated in the UK sample, location and device type were significant predictors of Relationship Dependency, with mobile and on-the-move use linked to higher scores - patterns not observed in the Arab sample.

Taken together, these results point to three important conclusions. First, the LLM-D12 is sufficiently sensitive to variations in LLM usage patterns, indicating its potential practical utility for practitioners. Second, the findings align with established research in technology design, where product designers aim to anticipate and meet user needs by examining intentions and experiences in order to enhance usability [59]. In our sample, for example, multimodal interaction was associated with higher instrumental dependency, suggesting that users value this modality for its perceived utility and accessibility. This insight could inform design decisions aimed at improving functionality and user satisfaction. However, it also highlights the need for caution, as features that enhance accessibility may inadvertently contribute to overreliance on LLMs. At the same time, the results indicate that the accessibility of LLMs may entail psychological costs, reflected in elevated levels of dependency. Third, while some usage-related predictors, such as interaction method, appear consistent across cultural contexts, others, including frequency of use, device type, and location, may be more culturally specific in their association with LLM dependency. These findings may be particularly relevant for professionals involved in the design, deployment, and regulation of AI systems, as they demonstrate the importance of accounting for both usage patterns and cultural context when assessing or managing user dependency on LLMs.

**Implications for Research and Practice**

LLM dependency is an emerging area of concern, yet there remains a global lack of validated instruments to assess this construct. This study presents the first validated instrument to assess LLM dependency in an Arabic-speaking population. In light of the growing body of literature documenting the potential negative consequences associated with the use of generative AI, and the increasing global demand for AI regulations [60] [61], there is a pressing need for reliable tools to measure users' dependency on such technologies. The validated Arabic version of the LLM-D12 directly addresses this need and facilitates further research into the psychological and social implications of LLM engagement within Arabic-speaking contexts.

The most immediate application of the Arabic LLM-D12 is to document empirical evidence of LLM dependency in the region. This foundational work is crucial for revealing the baseline of reliance of LMMs in this region and nature of LLM integration into daily life. Prior research has indicated generally positive attitudes toward AI among Arab populations; for example, studies have shown higher acceptance and lower fear compared to the UK [62] [56], raising the question of whether such attitudes extend to AI-LLMs and influence patterns of LLM dependency. Understanding these patterns is essential for anticipating how dependency may develop. The data generated from research using this tool can provide the empirical evidence needed to craft national and regional AI policies that are not only innovative but also ethical and human-centric. This aligns with global governance frameworks, such as the UNESCO Recommendation on the Ethics of AI [63], which has been the subject of regional consultations and calls for AI development that promotes human rights, well-being, and accountability.

A particularly pressing area for research is the relationship between LLM dependency and mental health. Surveys consistently show that Arab youth are highly engaged with digital technologies and are simultaneously concerned about the negative impact of technology addiction on their mental well-being

[64] [65] [66]. The validated LLM-D12 can be used in longitudinal studies along with measurements of other forms of behavioural problematic use such as social media, internet and gaming additions.

The scale's dual-dimensional structure allows for a more nuanced research program that can disentangle the distinct consequences of each type of dependency. For example, the Instrumental Dependency scale provides a direct means to test the ongoing debate about whether AI acts as an "intelligence amplifier" or a cognitive "crutch". Researchers can correlate high Instrumental Dependency scores with objective measures like academic performance and standardized tests of critical thinking. Drawing a parallel to prior findings in problematic digital media use research, including internet, social media, and gaming, where excessive engagement has been linked to lower academic outcomes [67] [68][69]. Such comparative framing may help determine whether AI dependency follows similar patterns to digital media overuse. The Relationship Dependency scale can facilitate a targeted investigation into the formation of parasocial bonds with AI. Research can also explore whether high Relationship Dependency scores are associated with reduced quantity or quality of face-to-face interactions, increased feelings of loneliness, or a preference for AI companionship due to its controllable and non-judgmental nature. This mirrors findings from research on digital media use where excessive engagement is often linked to has been linked to greater social withdrawal, heightened loneliness, and reduced in-person social interaction [70] [71] [72]. Such parallels underscore the need to explore whether AI dependency follows similar psychosocial patterns. This is particularly relevant in the context of the MENA "digital paradox," where a societal preference for relational technologies may predispose users to this form of dependency.

Research using the Arabic LLM-D12 can provide policymakers with the empirical data needed to craft more human-centric AI governance. Moreover, one of the most significant concerns voiced by educators is that over-reliance on AI will lead to "cognitive offloading," where students use technology as a substitute for, rather than a supplement to, their own intellectual labour [73] [74]. The LLM-D12 can be used to identify students at high risk for this behaviour, allowing for proactive pedagogical interventions.

**Limitations**

The validation of this scale should be considered within the context of several limitations. First, the cross-sectional design precludes any inference about the development or progression of LLM dependency over time. Future research should employ longitudinal designs to assess whether high dependency scores remain stable, increase, or decline over time, and whether they can predict the onset or persistence of problematic behaviours. It is also important to consider the potential role of a *novelty effect*, whereby individuals may initially show overuse, or increased productivity with a new technology, only for such patterns to diminish and stabilize as the novelty wears off. Tracking these trajectories could help to understand whether elevated dependency scores reflect transient adjustment to a new tool or a more enduring behavioural pattern. Moreover, exploring the unfolding dynamics of LLM dependency over time may provide insights into the psychosocial mechanisms underlying this form of behavioural dependency.

Second, the sampling strategy relied on voluntary participation through online platforms, a common approach in early-stage research on digital technologies. While this method is practical and efficient, particularly for accessing users already familiar with LLMs, it introduces self-selection bias and limits the generalisability of the findings. Respondents who actively engage with LLMs and choose to participate in such surveys may differ meaningfully from those who do not, especially in their attitudes towards AI, usage patterns, and digital literacy. However, this limitation is better understood as a deliberate scope condition of the study. Measuring LLM dependency meaningfully requires participants who already use these systems frequently. Accordingly, we recommend that the scale be applied primarily to regular or active LLM users, for whom the construct is most relevant. Furthermore, the relative demographic homogeneity of the sample may obscure within-group variability that is highly relevant in diverse Arab populations.

Third, the study did not incorporate qualitative data on participants' subjective experiences or interpretations of Instrumental and Relationship dependency. Such insights could have added depth to the construct validation process, particularly in highlighting culture-specific nuances in how dependency is perceived and enacted. While the current focus was on the psychometric development and validation of the LLM-D12, future research phases will integrate qualitative methodologies to support item refinement and broader conceptual clarity.

Fourth, the construct of LLM dependency is an emerging area of study. While the development of LLM-D12 scale was informed by a prior concept paper that provided a substantial theoretical foundation [10], ongoing research may reveal additional dimensions of the construct that warrant consideration. Terms such as *addiction*, *overuse*, *overreliance*, and *dependency* are often used interchangeably in literature, despite their distinct implications. We caution against interpreting LLM dependency as inherently pathological in the absence of robust evidence demonstrating negative impacts on well-being, cognitive performance, or social functioning. However, by operationalising dependency into measurable dimensions, this study provides a foundation for future empirical work to examine whether and under what conditions such reliance may become maladaptive.

Finally, although frequency of use was included as an external validator, a more comprehensive understanding of LLM engagement would benefit from additional behavioural metrics. Future work might examine variables such as duration of use, type of interaction, content generated, or specific use cases, each of which could offer more detailed insight into how dependency manifests in practice.

## CONCLUSION

This study evaluated the psychometric properties of the LLM-D12 questionnaire and validated it as a measure of LLM dependency in an Arabic-speaking population. The findings demonstrate that the scale is both valid and reliable for assessing two distinct dimensions of dependency: Instrumental and Relationship Dependency. Both subscales showed good internal consistency and were supported through external validation analyses. The scale also demonstrated sensitivity to actual LLM usage, indicating its applicability in both research and applied settings.

The validated LLM-D12 provides a practical tool for researchers examining the psychological and behavioural impact of generative AI, educators monitoring students' engagement with such tools, clinicians exploring emerging forms of technology-related distress, and policymakers developing evidence-based responses to AI adoption. It enables systematic investigation into the social and psychological effects of LLM use in the Arab context, where such tools are rapidly becoming part of everyday life.

**Acknowledgments**

This publication was supported by NPRP- 14-Cluster Grant # NPRP 14 C- 0916–210015 from the Qatar National Research Fund (a member of the Qatar Foundation). The findings herein reflect the work and are solely the authors' responsibility.

**Competing Interests**

The authors report no conflicts of interest.

**Data Availability Statement**

The dataset and analyses are available on the Open Science Framework (OSF) at the following link (DOI 10.17605/OSF.IO/Q9GZT) https://osf.io/q9gzt/

**Authors' contributions**

**SA**: Conceptualized and designed the research, translated the original questionnaire and curated data, assisted with statistical analysis, and wrote the first draft. **AY**: Conceptualized and designed the research, designed, performed and reported statistical analysis, and wrote the first draft. **ML**: Participated in the design of the original scale and critically reviewed and edited the paper. **RA**: Conceptualized, designed and coordinated the research, translated the original questionnaire and curated data, acquired research funding, and critically reviewed and edited the paper.

**Ethics statement and consent to participate**

Ethics approval for this study was obtained from Ethical approval for the study was obtained from Research Ethics Panel for the Science, Technology at Bournemouth University, UK (62239, 03/03/2025). Informed consent was obtained from all participants. All methods were performed in accordance with the relevant guidelines and regulations laid down in the Declaration of Helsinki.

**Appendix 1**

**Large Language Model Dependency Questionnaire (LLM-D12) – The English Version**

Thinking specifically about the large language model (LLM) you use most often (such as ChatGPT, DeepSeek, Gemini, Copilot, etc.), please indicate the extent to which you agree or disagree with each of the following statements:

| Item nr | item | Strongly Disagree | Disagree | Somewhat Disagree | Somewhat Agree | Agree | Strongly Agree |
|---|---|---|---|---|---|---|---|
| ID1 | Without it, I feel less confident when making decisions | | | | | | |
| ID2 | I use it sometimes without realizing how much time I spend immersed in it | | | | | | |
| ID3 | I feel much more rewarded and pleased when completing tasks using it | | | | | | |
| ID4 | I turn to it for support in decisions, even when I can make them myself with some effort | | | | | | |
| ID5 | It is my go-to for assistance in decision-making | | | | | | |
| ID6 | Making decisions without it feels somewhat uneasy | | | | | | |
| RD1 | I share details about my private life with it | | | | | | |
| RD2 | I interact with it as if it were a genuine companion | | | | | | |
| RD3 | It helps me feel less alone when I need to talk to someone | | | | | | |
| RD4 | It adds to my social life, making socializing more engaging and interesting | | | | | | |
| RD-5 [R] | I use it solely as a tool, not to express my feelings or expect it to understand me | | | | | | |
| RD6 | It helps me feel less alone, reducing the need to talk to others | | | | | | |

**Scoring:**

Items are rated on a 6-point scale: 1 = Strongly Disagree, 2 = Disagree, 3 = Somewhat Disagree, 4 = Somewhat Agree, 5 = Agree, 6 = Strongly Agree.

[R] indicates that the item shall be reversed when scoring

ID – Instrumental Dependency, RD – Relationship Dependency

Instrumental and Relationship Dependency should each be calculated as the sum of responses on their respective items. The two scales and their scores can be used separately. It is also possible to use the total score for LLM Dependency, obtained by summing the scores of both scales.

**Notes:**

The LLM examples in the introductory context can be updated to reflect either the most popular models at the time or those most relevant to the context in which the scale is administered.

The scale may be used at the item level, subject to several considerations and limitations. Please refer to Appendix 2 of the English version of the scale, accessible at: https://doi.org/10.48550/arXiv.2506.06874

# النسخة العربية من مقياس الاعتماد على النماذج اللُّغوية الكبيرة (LLM-D12)

فكَّر في النموذج اللُّغوي الكبير (LLM) الأساسيّ الخاص بك، أي النموذج الذي تستخدمه أكثر من غيره. (مثلا: ChatGPT, DeepSeek, Gemini, Copilot, إلخ).

يرجى تحديد مدى موافقتك أو معارضتك بما يتعلَّق بكل من العبارات التالية:

| رقم البند | البند | أعارض بشدة | أعارض | أعارض إلى حد ما | أوافق إلى حد ما | أوافق | أوافق بشدة |
|---|---|---|---|---|---|---|---|
| ID1 | من دونه، أشعر بثقة أقل عند اتخاذ القرارات | | | | | | |
| ID2 | أستخدمه أحيانًا دون أن أدرك كم من الوقت أقضيه مندمجًا معه | | | | | | |
| ID3 | أشعر بمزيد من الرضا والمتعة عند إكمال المهام باستخدامه | | | | | | |
| ID4 | ألجأ إليه لدعمي في اتخاذ القرارات، حتى عندما يمكنني اتخاذها بنفسي ببعض الجهد | | | | | | |
| ID5 | هو وجهتي المفضلة للحصول على المساعدة في اتخاذ القرارات | | | | | | |
| ID6 | اتخاذ القرارات بدونه يشعرني بنوع من عدم الارتياح | | | | | | |
| RD1 | أشارك معه تفاصيل عن حياتي الخاصة | | | | | | |
| RD2 | أتفاعل معه كما لو كان رفيقًا حقيقيًا | | | | | | |
| RD3 | يساعدني على تخفيف الشعور بالوحدة عندما أحتاج إلى التحدث مع أحد | | | | | | |
| RD4 | يضيف إلى حياتي الاجتماعية، مما يجعل التفاعل الاجتماعي أكثر متعة وإثارة | | | | | | |
| RD-5 [R] | لا أتوقع منه أن يفهمني ولا أشارك او أتناقش معه كيف أشعر ، استخدمه فقط كأداة | | | | | | |
| RD6 | يساعدني على الشعور بقدر أقل من الوحدة، مما يقلل حاجتي للتحدث مع الآخرين | | | | | | |

**التقييم:**

يتم تقييم البنود على مقياس من 6 نقاط: 1 = أعارض بشدة، 2 = أعارض، 3 = أعارض إلى حد ما، 4 = أوافق إلى حد ما، 5 = أوافق، 6 = أوافق بشدة.

[R] تشير إلى أنه يجب عكس الإجابة عند التقييم

**ID** = الاعتماد الوظيفي ، **RD** = الاعتماد العلائقي

تُحسب كلٌّ من درجة الاعتماد الوظيفي والاعتماد العلائقي على حدة من خلال مجموع الإجابات على البنود الخاصّة بكل منهما. يمكن استخدام هذين المقياسين الفرعيين ودرجاتهما بشكل منفصل كما يمكن استخدام الدرجة الإجمالية للاعتماد على النماذج اللغوية الكبيرة والحصول عليها من خلال جمع درجتيّ المقياسين

**ملاحظات:**

يمكن تحديث أمثلة نماذج اللغة الكبيرة (LLMs) في السياق التمهيدي لتعكس إمّا النماذج الشائعة في ذلك الوقت أو النماذج الأكثر صلة بالسياق الذي يُطبَّق فيه المقياس.

يمكن أيضًا استخدام المقياس على مستوى البنود، مع مراعاة عدة اعتبارات وقيود. يُرجى الرجوع إلى الملحق 2 من النسخة الإنجليزية للمقياس، والمتاح عبر الرابط:
https://doi.org/10.48550/arXiv.2506.06874